\documentclass[10pt,twocolumn,letterpaper]{article}

\usepackage{cvm}
\usepackage{times}
\usepackage{epsfig}
\usepackage{graphicx}
\usepackage{amsmath}
\usepackage{amssymb}
\usepackage[vlined,ruled,norelsize,commentsnumbered,linesnumbered]{algorithm2e}

\usepackage[pagebackref=true,breaklinks=true,letterpaper=true,colorlinks,bookmarks=false]{hyperref}

\makeatletter
\newcommand{\removelatexerror}{\let\@latex@error\@gobble}
\makeatother

\cvmfinalcopy 

\def\cvmPaperID{22} 

\ifcvmfinal\fi
\begin{document}

\title{Light Transport Simulation via Generalized Multiple Importance Sampling}

\author{Qi Liu\\
Shanghai Jiao Tong University\\
\and Yiheng Zhang\\
Shanghai Jiao Tong University
\and Lizhuang Ma\\
Shanghai Jiao Tong University
\\
}

\maketitle

\begin{figure*}
	\begin{center}
		\includegraphics[width=\textwidth]{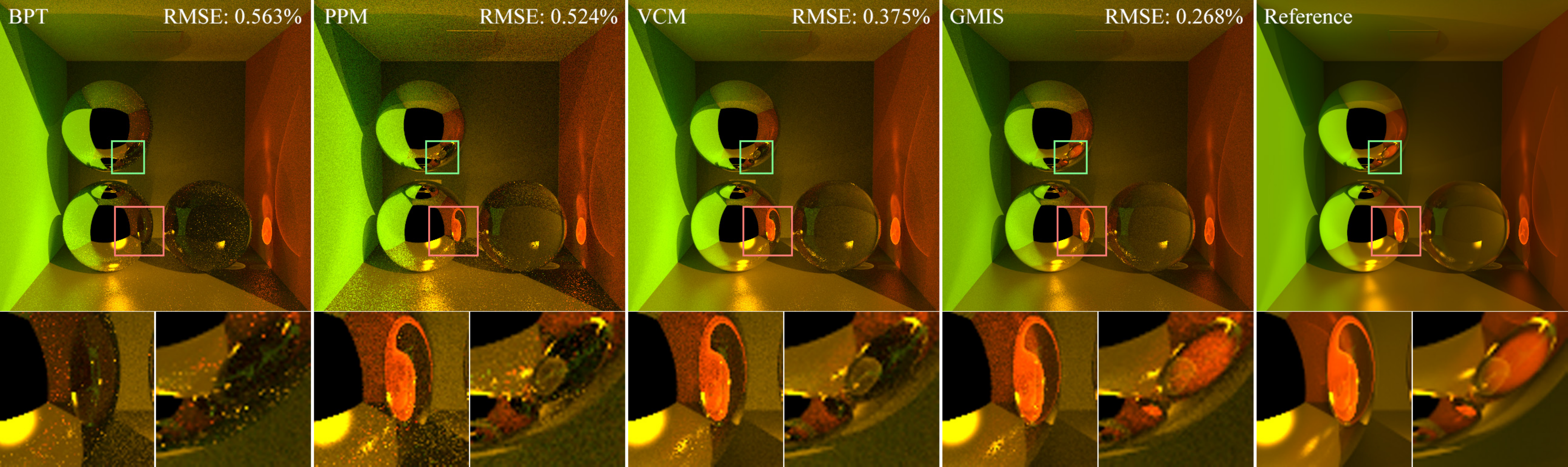}
		\caption{Comparison of our generalized multiple importance sampling (GMIS) algorithm with bidirectional path tracing (BPT), Progressive photon mapping (PPM), and vertex connecting and merging (VCM) after 30 minutes of rendering. GMIS show that better convergence than other methods, as it could capture illumination of entire scenes in more efficient ways. Results show that GMIS catches more details in specular shadows, while in glossy areas the color is more smooth than previous methods.}
	\end{center}
\end{figure*}

\begin{abstract}
    Multiple importance sampling (MIS) is employed to reduce variance of estimators, but when sampling and weighting are not suitable to the integrand, the estimators would have extra variance. Therefore, robust light transport simulation algorithms based on Monte Carlo sampling for different types of scenes are still uncompleted. In this paper, we address this problem by present a general method, named generalized multiple importance sampling (GMIS), to enhance the robustness of light transport simulation based on MIS. GMIS combines different sampling techniques and weighting functions, extending MIS to a more generalized framework. Meanwhile, we implement the GMIS in common renderers and illustrate how it increase the robustness of light transport simulation. Experiments show that, by applying GMIS, we obtain better convergence performance and lower variance, and increase the rendering of ambient light and specular shadow effects apparently.
\end{abstract}

\section{Introduction}

Monte Carlo light transport simulation is an important technique in photorealistic rendering. Existing simulation algorithms often work well, but they still face the convergence problem. Multiple importance sampling (MIS) is introduced to alleviate this problem, it tries to lower the impact of the variance by combining several estimators, each using one strategy that correctly matches one part of the integrand, but different estimators may have apparent variance for various types of scenes. Thus, designing estimators with lower variance and developing robust solutions under different scene configurations should be put more attention.

MIS could not give information about which sampling strategy should be preferably used to lower variance for a given integrand. Since MIS could minimize variance of estimators, but if sampling configuration is less adapted to different scenes, the estimator would suffer from extra variance. Existing methods on MIS\cite{Pajot:2011:RRA:1999161.1999196} addressed this issue by introducing representativity for adaptive choosing among the different sampling strategies, including BSDF-based, photon-map-based strategy representativity and other sampling configurations. It could find optimal sampling configurations for different integrands that require different sampling configurations, obtaining an optimal estimator with respect to variance. VCM/UPS~\cite{Georgiev:2012:VCM, Hachisuka:2012:PSE:2366145.2366210} followed similar idea, presenting an integration of bidirectional path tracing and photon mapping into a framework using vertex connecting and merging. It employed bidirectional scattering distribution function (BSDF)\cite{article} information aligned with photon mapping (PM)\cite{Jensen:2001:RIS:500844} to compute an extended MIS. These methods utilized extra assistant information to calculate MIS to improve quality of light transport, while considering less robust numerical integration for lower variance. 

In this paper, we propose a more robust numerical integration framework for better convergence and lower variance. We combine three sampling strategies and five weighting functions, obtaining a more generalized MIS framework, named GMIS. GMIS impose the sampling quality both on specular and non-specular event, it could increase sampling numbers on non-specular events, while it also extend path length on specular surface. We also demonstrate the implementation of the GMIS into existing light transport algorithms. Finally, we obtain more realistic rendering effects with fast convergence rate, as color around ambient light area is more smooth and specular surfaces and their specular shadows contain more details with less noises. Experiments also show that GMIS has the lowest variance compared with other methods.

The rest of the paper is organized as follows. In section 2, we list the previous work of light transport simulation, while the background knowledge of existing MIS methods will be described in section 3. In section 4, we are going to explain the theory of the GMIS. In section 5, we talk about how to implement the GMIS into existing renderer. Then, we test various performance of the GMIS to illustrate its robustness. Finally, we will summarize the paper and make some conclusions.

\section{Previous Work}
This section reviews several famous method of global illumination: path tracing, photon mapping and metropolis light transport. Developments and strengths are analyzed respectively.
\subsection{Path tracing}
Path tracing (PT) algorithm has long been considered as a natural way to simulating the transport of light. By generating random lights from camera, the naive path tracing algorithm \cite{Kajiya:1986:RE:15886.15902} utilize physical characteristics of light transportation to render the final image for one scene. It was proved later, with another way round, tracing light basically from light sources \cite{Dutré1995} could improve the efficiency of path tracing algorithm and thus enhance the quality of final result. Bidirectional path tracing \cite{Lafortune93bi-directionalpath, Veach1995} was invented to further accelerating the speed of path tracing algorithm by sampling from the camera and light sources at the same time, counting more contribution to each pixel. With the introduction of multiple importance sampling scheme to traditional rendering process, a robust structure of path tracing was attained in 1995 \cite{Veach:1995:OCS:218380.218498}, which brought path tracing algorithm to a rather mature level. Traditional path tracing methods are not used nowadays, but they are still the bases of many advanced global illumination methods. Our work is to seek the possibility of optimize the multiple importance sampling process, with a further discussing in sampling proposals as well as the method of evaluating weight function. As in the path tracing part in our algorithm, we present a new way of sampling the light scattering process.
\subsection{Photon mapping}
Photon mapping (PM) algorithm\cite{Jensen:2001:RIS:500844} focus on estimating radiance of different location of the scene and reuses the radiance value stored in cache during scattering photons in the light path. Compare to path tracing algorithms, photon mapping outperforms especially in the specular-diffuse-specular condition, since it provides a more precise estimation by combining nearby photons in the photon map. Progressive photon mapping (PPM)\cite{Hachisuka:2008:PPM:1409060.1409083} provides a new method of calculating the radiance of photon mapping, thus excludes the bias computation in the traditional photon mapping algorithm. Stochastic progressive photon mapping \cite{Hachisuka:2009:SPP:1618452.1618487} computes the correct average radiance value over a region by a simple algorithmic modification of PPM. 
Vertex connection and merging(VCM) algorithm \cite{Georgiev:2012:VCM} combined the multiple importance sampling estimator in path tracing and photon mapping into a more general framework. With the formulation of recursive function in the computation of multiple importance sampling, VCM dramatically increase the speed of integral calculation. Our method managed to optimize the MIS estimator through a new way of calculating the samples' weight function.
\subsection{Metropolis light transport}
Metropolis light transport (MLT) algorithm \cite{Veach:1997:MLT:258734.258775} introduces Markov Chain Monte Carlo (MCMC) method to the rendering problem. With the random number sequence, MLT generate light path in a non-physical way, taking the advantage of heuristic algorithms. The key idea is to apply mutation to initial light paths and rapidly produces subpaths, which significantly increase the performance of renderer when dealing with specular-diffuse-specular light. However, the MCMC methods requires an effective mutation function of the sampler to ensure the fast convergence of final result. Primary sample space MLT (PSSMLT)\cite{CGF:CGF703} simplifies the generation of MCMC, which defines the path $\overline{x}$ by a vector of random numbers. Multiplexed metropolis light transport (MMLT) algorithm \cite{Hachisuka:2014:MML:2601097.2601138} further improves the efficiency and robustness by taking previous MIS factor into account during the process of sampling Markov Chain with a easy implementation.
\subsection{Learning methods}
Recent year, various learning methods are introduced to the light transport simulation. Progressive Gaussian mixture model (GMM)\cite{Vorba:2014:OLP:2601097.2601203} is presented to calculate spatial scattering of scalar radiance in participating media, which is good fit to the initial particles. Subsequent work extended\cite{CGF:CGF12950} GMM to a product importance sampling, finding a good approximation to the illumination integrand as a sampling distribution during rendering. For path guiding, Thomas et al.\cite{10.1111:cgf.13227} introduced SD-trees to guide light paths for high-energy sampling, and Ken et al.\cite{DBLP:journals/corr/DahmK17} employed Q-learning to sample light transport paths to visibility-important areas. In deep learning, Nima et al.\cite{Kalantari:2015:MLA:2809654.2766977} used multilayer perceptron (MLP) to de-noise rendering images. Oliver et al.\cite{Nalbach:2017:DSC:3128450.3128458} introduce convolutional neural network (CNN), learning how to transfer several noisy images into a smooth image like reference. These learning methods could provide better rendering images, but the details contain much artifacts, as the neural network methods are more like image enhancement. Thus, the result images are not photorealistic in many areas, such as edges around sharp objects.

\section{Background}

\subsection{Multiple importance sampling}
As a universally acknowledged method, Monte Carlo method is a deterministic way of evaluating a specific integral by sampling $N$ samplers $x_i \in \Omega$ from a given probability density function (pdf) $p(x)$ as
\begin{equation}
I = \int_{\Omega}{f(x)dx}\approx \frac{1}{N} \sum_{i=1}^{N}\frac{f(x_i)}{p(x_i)}
\end{equation}
To increase the robustness of the estimation process, multiple importance sampling(MIS)\cite{} uses $M$ proposal functions together with corresponding weight value to estimate the target integral as
\begin{equation}
\langle I\rangle = \sum_{i=1}^{M}\frac{1}{n_i}\sum_{k=1}^{n_i}w_i(x_{i,k})\frac{f(x_{i,k})}{p(x_{i,k})}
\end{equation}
where $x_{i,k}$ are independent variables with proposal functions $p(x_{i,k})$, and $w_i(x_{i,k})$ serves as the weight function of each sample.

\subsection{Path integral}
The path integral\cite{Veach:1998:RMC:927297} forms the key measurement of light transport, which evaluates the color of each pixel in the final image as \begin{equation}
I = \int_{\Omega}{f(\overline{x})d\mu(\overline{x})}
\end{equation}
where $\overline{x} = x_0,x_1,\dotsm x_j$ denotes an independent light path with $j$ edges and $j+1$ light vertices. Here, $x_0$ denotes the light source while $x_j$ denotes the ending point in the camera. $f$ is the contribution measurement function and $\mu$ is the area measure with $\Omega$ being the light path space.

To further analyze the contribution function. $f(\overline{x})$ can be rewritten as bellow:
\begin{equation}
f(\overline{x}) = L_e(x_0)G(x_0,x_1)S(\overline{x})W_e(x_j)
\end{equation}
In this equation, $L_e$ denotes the emission radiance from given light source. $G$ stands for the geometry term in light transport and $S$ is the scattering factor which measures the contribution of each light vertex except the light source and camera point in the light path. $W_e$ represents the sensitivity of specific pixel, in other words, the ending point in the camera. Specially, $S$ could be expanded in a multiple multiplication form:
\begin{equation}
	S(\overline{x}) = \prod_{i=1}^{j-1}f_s(x_i)G(x_i,x_{i+1})
\end{equation}
where $f_s$ is the bidirectional scattering factor(BSDF) at a given surface.

\subsection{Contribution estimation}
The vertex connection and merging algorithm (VCM)\cite{Georgiev:2012:VCM} separate the computation of contribution $I$ to a given pixel by two parts, $I_{VC}$ and $I_{VM}$, each of which is estimated with MIS method.
\begin{equation}\label{equ:vcm}
	\langle I\rangle_{\mathbf{VCM}} = \frac{1}{n_{\mathbf{VC}}}\sum_{i = 1}^{n_{\mathbf{VC}}}\langle I_i\rangle_{\mathbf{VC}} + \frac{1}{n_{\mathbf{VM}}}\sum_{i = 1}^{n_{\mathbf{VM}}}\langle I_i\rangle_{\mathbf{VM}}
\end{equation}
\begin{equation}
\langle I_i\rangle_{\mathbf{VC}} =
\sum_{k=1}^{n_i} w_{i,\mathbf{VC}} (\overline{x_k})\frac{f_{\mathbf{VC}}(\overline{x_k})}{p_{\mathbf{VC}}(\overline{x_k})}
\end{equation}
\begin{equation}
\langle I_i\rangle_{\mathbf{VM}} =
\sum_{k=1}^{n_i} w_{i,\mathbf{VM}} (\overline{x_k})\frac{f_{\mathbf{VM}}(\overline{x_k})}{p_{\mathbf{VM}}(\overline{x_k})}
\end{equation}

Furthermore, VCM formulate a recursive expression for MIS weight factors both in sampling the vertex connection and vertex merging, which enables the estimator to reuse previous results, significantly increasing the overall efficiency. The vertex connection and vertex merging weight factor can be expressed as bellow.
\begin{equation}
w_{0,\mathbf{VC}} = \frac{\stackrel{\leftarrow}{p_0}}{\stackrel{\rightarrow}{p_0}}
\end{equation}
\begin{equation}
w_{i,\mathbf{VC}} =  \stackrel{\leftarrow}{p_i}(\eta_{\mathbf{VCM}}+\frac{1}{\stackrel{\rightarrow}{p_i}}+\frac{1}{\stackrel{\rightarrow}{p_i}}w_{i-1,\mathbf{VC}})
\end{equation}
\begin{equation}
w_{1,\mathbf{VM}} = \stackrel{\rightarrow}{p_1}(\frac{1}{\eta_{\mathbf{VCM}}}+\stackrel{\leftarrow}{p_0}\frac{1}{\eta_{\mathbf{VCM}}\stackrel{\rightarrow}{p_0}})
\end{equation}
\begin{equation}
w_{i,\mathbf{VM}} =  \stackrel{\rightarrow}{p_i}(\frac{1}{\eta_{\mathbf{VCM}}}+\stackrel{\leftarrow}{p_{i-1}}+\stackrel{\leftarrow}{p_{i-1}}w_{i-1,\mathbf{VM}})
\end{equation}
Here, $\eta_{\mathbf{VCM}} = \frac{n_{\mathbf{VM}}}{n_{\mathbf{VC}}}\pi r^2$ in which $r$ is the vertex merging radius. $\stackrel{\rightarrow}{p_{i}}$ denotes the forward pdf of given vertex while $\stackrel{\leftarrow}{p_{i}}$ denotes the reverse pdf.

We take the advantage of this efficient method into our MIS weight factor calculation. But using another form of weight factor to enhance the robustness of our algorithm and smooth the final image.

\section{Generalized Multiple Importance Sampling}

In this section, we are inspired by Victor's work\cite{2015arXiv151103095E}. Multiple importance sampling (MIS) schemes consider \emph{N} proposal probability density functions (pdfs), $ \left \{ p_{1}(x), p_{2}(x),...,p_{N}(x)) \right \} $, and generate \emph{M ($ M\geq N $)} samples from pdfs with proper weight. Thus, the process of MIS contains two step: generating samples from existing pdfs, and evaluating weighting for each sample. In the following, we show all possible sampling techniques and weighting functions, and demonstrate different schemes by their combinations.

\begin{figure}\label{fig_sample}
	\begin{center}
		\includegraphics[width=\linewidth]{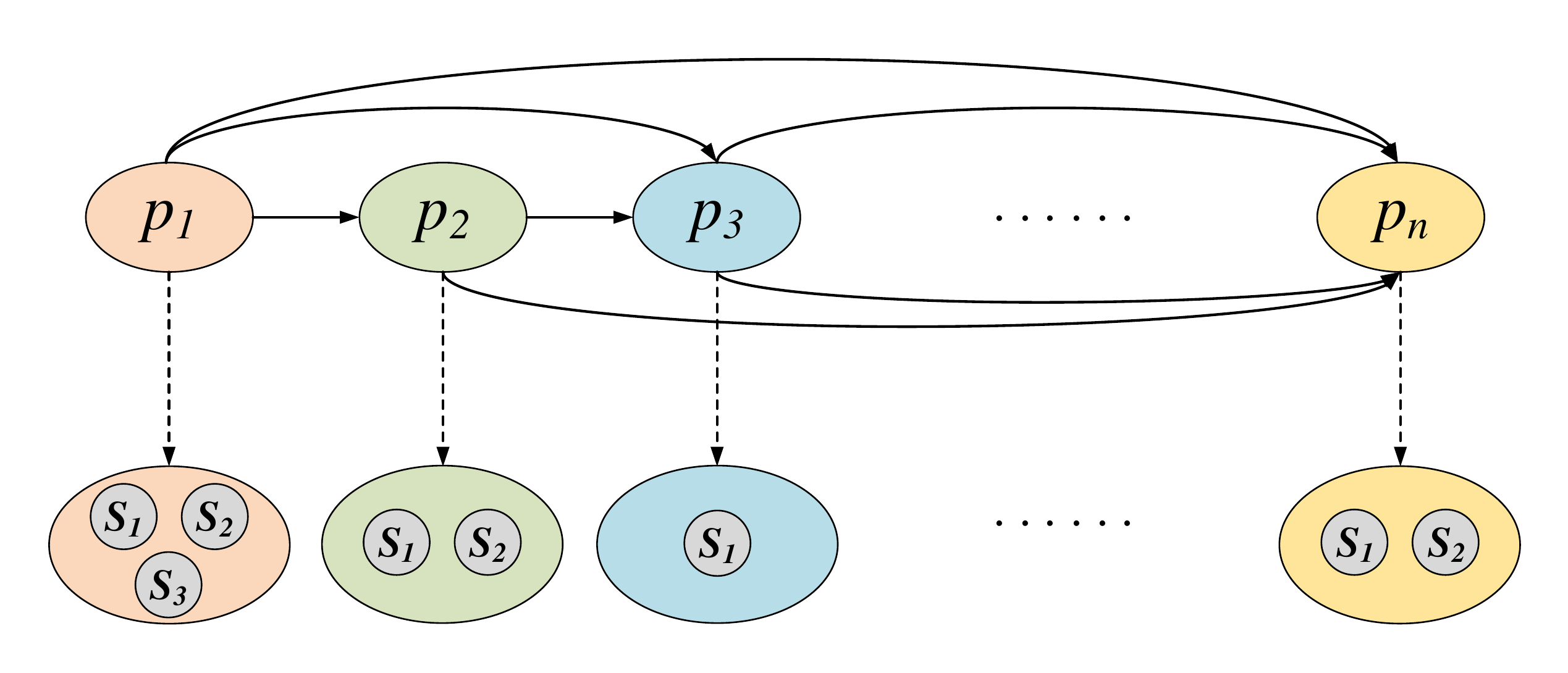}
		\caption{Sampling methodology. In this process, each probability density function (pdf) $p_{i}$ is generated based on the all previous pdfs. In each pdf, different number of samples could be obtained due to specific conditions.}
	\end{center}
\end{figure}

\subsection{Generating Samples}

For generating samples, it contains two steps: first is selecting the pdf to get a mixture of pdfs, and second is sampling from a mixture of pdfs.

Generally, the complete set of \emph{N} proposal pdfs, $ \left \{ p_{1}(x), p_{2}(x),...,p_{N}(x)) \right \} $, can be interpreted as a mixture of pdfs, 
\begin{equation}\label{equ:psi}
\psi \left ( x \right )\equiv \frac{1}{N} \sum_{n=1}^{N}p_{n}\left ( x \right ).
\end{equation}

First, we describe how to select pdfs. Here, there are 3 basic ways to generate from mixture pdfs:

\emph{S1}: random index selection with replacement.

\emph{S2}: random index selection without replacement.

\emph{S3}: deterministic index selection with replacement.

All these 3 selection mechanisms have same property that 
\begin{equation}
\frac{1}{N}\sum_{i=1}^{N}P(j_{i}=k)=\frac{1}{N},\quad \forall k\in \left \{ 1,...,N \right \},
\end{equation}
which means all the pdfs could be selected with same probability. Here, $ j_{i} $ denotes the index of proposal pdfs.

The next step is sampling from a mixture of pdfs. As shown in Figure \textbf{2}, generating samples $ \left \{ x_{1},...,x_{M} \right \} $ from the mixture pdf $ \psi $ is a sequential procedure. First, the \emph{n}-th index $ j_{n} $ is drawn from conditional pdf, $ P\left ( j_{n}\mid j_{1:n-1} \right ) $, where $ j_{1:n-1}\equiv \left \{ j_{1},...,j_{n-1} \right \} $ is the sequence of the previously generated indexes. Then, the \emph{n}-th sample is drawn from the selected proposal pdf as $ x_{m}\sim p\left ( x_{m}\mid j_{n} \right ) $. Generally, in theoretical proofs and implementations, the number of generated samples coincides with the number of proposal pdfs, i.e., $ M=N $. Here, we consider the extension case, $ M=kN $, with $ k\geq 2 $ and $ k\in N $ as shown in Fig.

\subsection{Evaluating Weights}

The weight is to evaluate the adequacy of the samples generated from pdfs with respect to target function. The weight assigned to the \emph{n}-th sample is proportional to the ratio between the target pdf and the pdf evaluated at each sample value,
\begin{equation}
\omega _{n}=\frac{f\left ( x_{n} \right )}{p\left ( x_{n} \right )}.
\end{equation}

The expectation of the generic estimator can be then computed as
\begin{equation}
E\left [ \hat{I} \right ]=\frac{1}{NZ}\sum_{n=1}^{N}\sum_{j_{1:N}}^{ }\int \frac{f(x_{n})g(x_{n})}{p(x_{n})}P(j_{1:N})p(x_{n}\mid j_{n})dx_{n}
\end{equation}

In the proposed framework, we consider valid any weighting scheme that yields $ E\left [ \hat{I} \right ]\equiv I $ in Eq.(x). We present 5 possible functions $ p(x_{n}) $ and different choices for $ p(x_{n}) $ come naturally from the sampling densities discussed in the previous section.

\emph{W1}: $ \omega(x_{n})=\psi _{j_{1:n-1}}(x_{n})=p(x_{n}\mid j_{1:n-1}) $

It interprets that the weight is the conditional density of $ x_{n} $ and all previous indexes of pdfs.

\emph{W2}: $ \omega(x_{n})=\psi _{j_{n}}(x_{n})=p(x_{n}\mid j_{n})=q_{j_{n}}(x_{n}) $

It interprets that if the $ j_{n} $ is known, the weight is the pdf with index $ q_{j_{n}}$.

\emph{W3}: $ \omega(x_{n})=p(x_{n}) $

It interprets that $ x_{n} $ does not need prior knowledge, only use existing pdf to calculate the weights.

\emph{W4}: $ \omega(x_{n})=\psi _{j_{1:N}}(x_{n})=\frac{1}{N}\sum_{k=1}^{N}q_{j_{k}}(x_{n}) $

It interprets that the weight is the distribution of $ x_{n} $ on the whole set of \emph{N} proposal pdfs.

\emph{W5}: $ \omega(x_{n})=\psi (x_{n})=\frac{1}{N}\sum_{k=1}^{N}q_{k}(x_{n}) $

It interprets that the weight is calculated based on all knowledge of $ p(x_{n})$.

\begin{figure}
	\begin{center}
		\includegraphics[width=\linewidth]{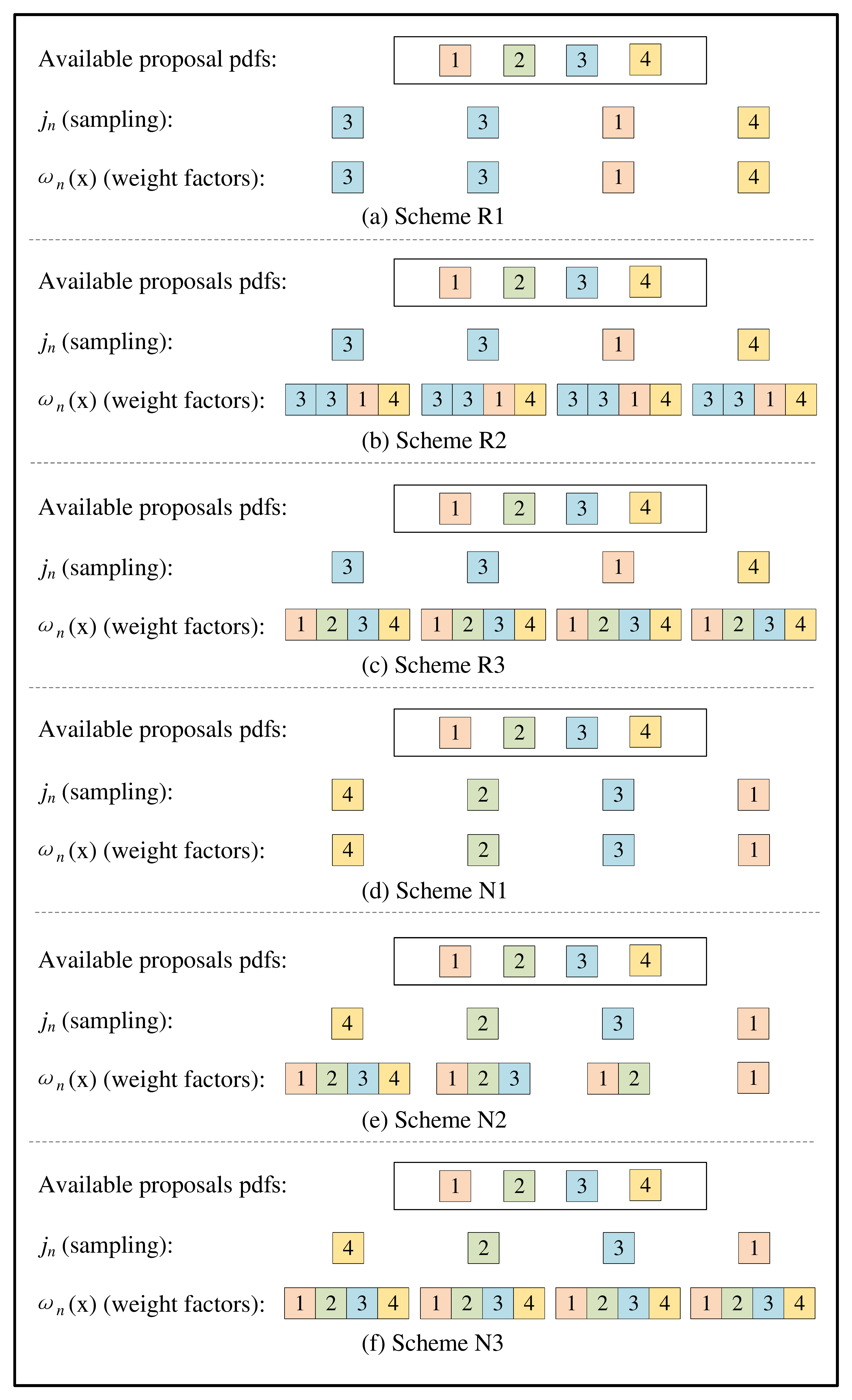}
		\caption{Six MIS schemes. The top three schemes are sampling with replacement, while the bottom three schemes select samples without replacement. For weight calculation, three different ways are coincide with these schemes.}
	\end{center}
\end{figure}

\subsection{Combination}

Since we have various sampling strategies and the weighting functions, we continue to describe the different possible combinations of them. We note that, even though we have discussed three sampling procedures and five alternatives for weight calculation, once combined the fifteen possibilities only lead to six unique MIS methods shown in Table \textbf{1}. Three of the methods are based on replacement \emph{S1}, while other three methods are associated to schemes without replacement \emph{S2, S3}.

\begin{table}[!hbp]
	\begin{center}
	\begin{tabular}{|c|c|c|c|c|c|}
		\hline
		   & \emph{W1} & \emph{W2} & \emph{W3} & \emph{W4} & \emph{W5} \\
		\hline
		\emph{S1} & \emph{R3} & \emph{R1} & \emph{R3} & \emph{R2} & \emph{R3} \\
		\hline
		\emph{S2} & \emph{N2} & \emph{N1} & \emph{N3} & \emph{N3} & \emph{N3} \\
		\hline
	    \emph{S3} & \emph{N1} & \emph{N1} & \emph{N1} & \emph{N3} & \emph{N3} \\
	    \hline
	\end{tabular}
	\end{center}
	\caption{Combination of the different sampling schemes and weighting functions.}
\end{table} 

For these six schemes, all \emph{R} schemes are sampling with replacement while all \emph{N} schemes are sampling without replacement. The main difference \emph{N} schemes and \emph{R} schemes is that whether they could sample a pdf same as previous selected pdfs. 
\emph{R} schemes contains \emph{R1, R2,} and \emph{R3} methods, they differ from each other in weight evaluation. For \emph{R1}, the weight is calculated by selected sample with existing pdf, and the weight in \emph{R2} is a posteriori mixture of a sequence of selected pdfs, while the weight in \emph{R3} is the whole mixture $\psi$ composed of all initial pdfs. The weight evaluation in \emph{N} schemes are same as corresponding \emph{R} schemes.

Since the convergence performance is important, we also demonstrate the variance of these six MIS schemes. The following equation is the variance analysis of six MIS schemes:
\begin{equation}
Var(I_{R1})=\frac{1}{N^{2}Z^{2}}\sum_{k=1}^{N}\int \frac{\pi^{2}(x)g^{2}(x)}{q_{k}(x)}dx-\frac{I^{2}}{N}
\end{equation}

\begin{equation}
\begin{split}
Var(I_{R2})=\frac{1}{N^{2}Z^{2}}\frac{1}{N^{N}}\sum_{j_{1:N}}^{ }\int \frac{\pi^{2}(x)g^{2}(x)}{f(x\mid j_{1:N})}dx\\-\frac{1}{N^{2}Z^{2}}\frac{1}{N^{N}}\sum_{j_{1:N}}^{ }\sum_{n=1}^{N}(\int \frac{\pi(x_{n})g(x_{n})}{f(x_{n}\mid j_{1:N})}q_{j_{n}}(x_{n})dx_{n})^{2}
\end{split}
\end{equation}

\begin{equation}
Var(I_{R3})=\frac{1}{NZ^{2}}\sum_{k=1}^{N}\int \frac{\pi^{2}(x)g^{2}(x)}{\psi(x)}dx-\frac{I^{2}}{N}
\end{equation}

\begin{equation}
Var(I_{N1})=\frac{1}{N^{2}Z^{2}}\sum_{n=1}^{N}\int \frac{\pi^{2}(x_{n})g^{2}(x_{n})}{q_{n}(x_{n})}dx_{n}-\frac{I^{2}}{N}
\end{equation}

\begin{equation}
\begin{split}
Var(I_{N2})=\frac{1}{N^{2}Z^{2}}\sum_{n=1}^{N}\sum_{j_{1:n-1}}^{ }\int \frac{\pi^{2}(x_{n})g^{2}(x_{n})}{p(x_{n}\mid j_{1:n-1})}P(j_{1:n-1})dx_{n}\\-\frac{1}{N^{2}Z^{2}}\sum_{n=1}^{N}\sum_{j_{1:N}}^{ }(\int \frac{\pi(x_{n})g(x_{n})}{p(x_{n}\mid j_{1:n-1})}q_{j_{n}}dx_{n})^{2}P(j_{1:n})
\end{split}
\end{equation}

\begin{equation}
\begin{split}
Var(I_{N3})=\frac{1}{NZ^{2}}\sum_{k=1}^{N}\int \frac{\pi^{2}(x)g^{2}(x)}{\psi(x)}dx\\-\frac{1}{N^{2}Z^{2}}\sum_{n=1}^{N}(\int \frac{\pi(x)g(x)}{\psi(x)}q_{n}(x)dx)^2
\end{split}
\end{equation}

Thus, we can get the relation between the variance of these six schemes. And the scheme \emph{N3} has the lowest variance among all MIS schemes.

\begin{equation}
Var(I_{R1})=Var(I_{N1})\geq Var(I_{R3})\geq Var(I_{N3})
\end{equation} 

\begin{equation}
\begin{split}
Var(I_{R1})=Var(I_{N1})\geq Var(I_{R2})=Var(I_{N2})\\ \geq Var(I_{N3})
\end{split}
\end{equation} 

\section{Implementation}

We implement our GMIS framework based on VCM~\cite{Georgiev:2012:VCM} as shown in Algorithm \textbf{1}, as VCM integrate various MIS techniques, combining bidirectional path tracing with photon mapping, while maintain the performance of bidirectional path tracing. In the render process, VCM first traces particles from light sources to build photon maps, and connect each particle directly to camera, obtaining an initial color value. Then, it traces a number of camera paths to estimate the value of each pixel, i.e., performing an estimation of the equation Eq.(\ref{equ:vcm}) using MIS at each point of camera path. The main 

In GMIS integration process, we focus on the building of photon maps by combining with \emph{N3} MIS scheme, as \emph{N3} could reach the lowest variance in all MIS schemes. When tracing photons emitted by light sources, existing methods only generate one next light event. Even VCM combines different factors that influence sampling, it is essentially a association of various importance factors, such as BSDF and photon maps. But this association of different importance factors is only multiple importance, it does not include multiple sampling. So, the implementation of \emph{N3} MIS scheme into current light transport simulation algorithms is indeed a generalized framework.

\begin{figure}[!t]\label{algorithm}
	\removelatexerror
	\begin{algorithm}[H]
		\BlankLine
		\SetKwInOut{Input}{input}
		\SetKwInOut{Output}{output}
		\caption{Implementation of GMIS}
		
		\For{i in PixelCount}
		{
			\emph{SampleNumber} = 0\;
			\emph{lightVertex} = TraceRay(CurrentPixel(i))\;
			\While{$SampleNumber\leq MAX$}
			{
				\If{hit on non-specular surface}
				{
					samples = SamplingForMultipleTimes()\;
					weightOfSamples = MixturePdfs()\;
				}
				\Else 
				{
					samples = SamplingForOneTimes()\;
					weightOfSamples = MixturePdfs()\;
				}
				MixtureWeightOfSamples()\;
				QueueAdd(samples)\;
				\emph{colorOfCurrentPixel} += ConnectToEye(samples)\;
				ContinueRandomWalk(samples)\;
				\emph{SampleNumber} += samples.size()\;
			}
		}
		BuildSpatialSearchStructure(lightVertices)\;
		
		\For{i in PixelCount}
		{
			\emph{eyeVertex} = TraceRay(CurrentPixel(i))\;
			\emph{colorOfCurrentPixel} += ConnectToLightSource(eyeVertex)\;
			\emph{colorOfCurrentPixel} += VertexConnecting()\;
			\emph{colorOfCurrentPixel} += VertexMerging()\;
			ContinueRandomWalk(eyeVertex)\;
		}
	\end{algorithm}
\end{figure}

First, to satisfy the sampling without replacement, the number of samples at each intersection point should be increased. Thus, a queue is introduced to save the unprocessed light vertex. Particularly, if a light hits on a diffuse surface, the number of sampling would be increased to get more smooth color. If the intersection is on a specular surface, generating reflect or refract event, the number of sampling would be reduced to one sample. For each path, we have a maximum number of samples, which means, all paths have equivalent samples. Thus, for a path with more specular event, the path length would be extended to capture more specular effects, while for a path contains more diffuse event, the length would be short with more samples in several diffuse intersection points, thus lead to more smooth color.

Then, when we get several samples in a light intersection point, we employ weighting function of 
\emph{N3} scheme to put the contribution values of each sample together. Thus, a MIS weight array is set to store the weight of each sample. When we calculate the final color of a intersection point, we assign the average weight, evaluated by Eq.(\ref{equ:psi}), to the weight value of each sample. The main benefit is, by a mixture of all different samples, we get a more smooth color in many areas.

\begin{figure*}\label{res}
	\begin{center}
		\includegraphics[width=\textwidth]{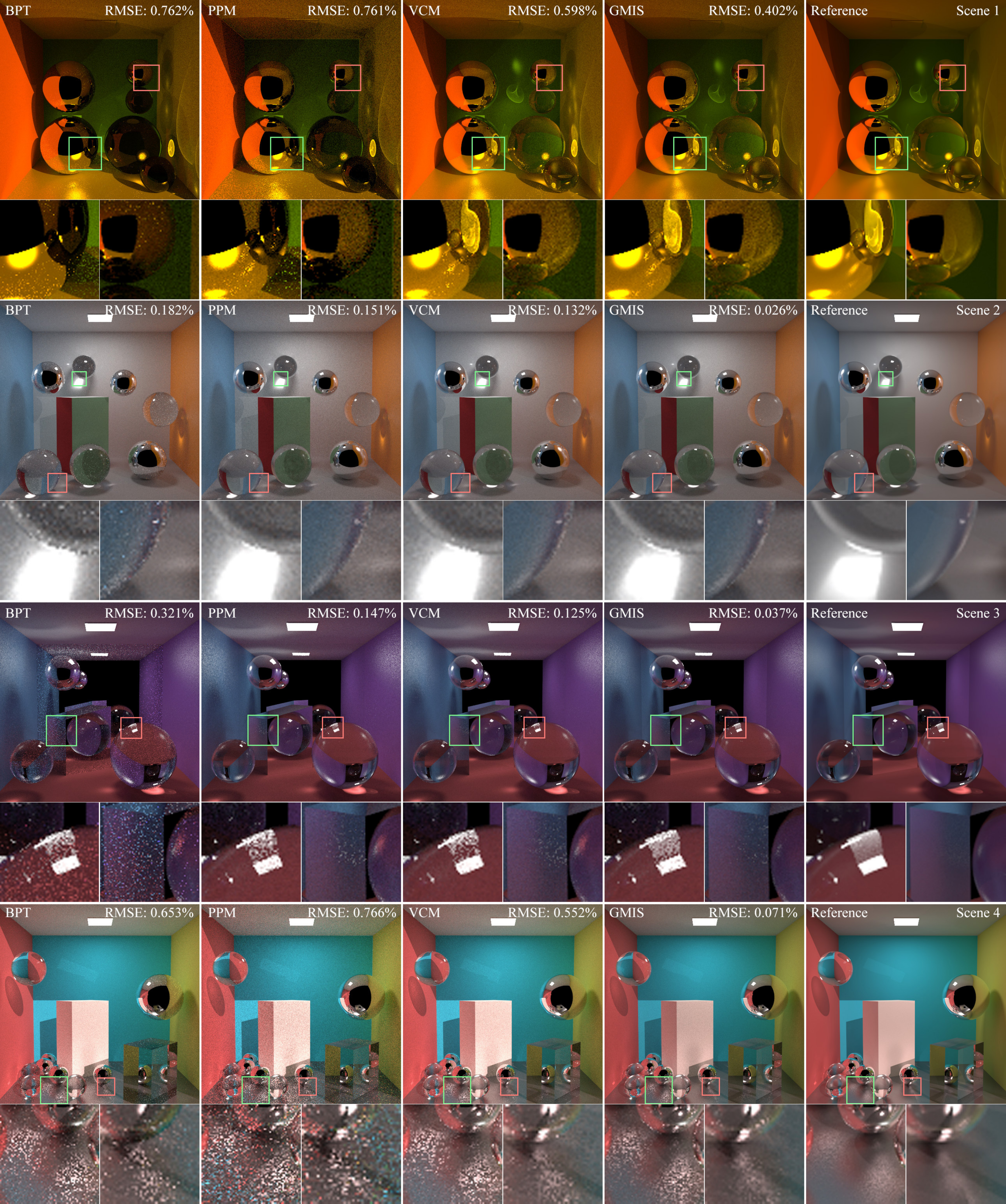}
		\caption{Results comparison of our generalized multiple importance sampling (GMIS) algorithm with bidirectional path tracing (BPT), progressive photon mapping (PPM) and vertex connecting and merging (VCM) after 5 minutes of rendering. The reference images on the right column have been rendered more than 24 hours. These four test scenes have different characters, including various types of diffuse, glossy, specular and transport surface, thus bring out complex light intersections. Our GMIS employ a more generalized sampling framework to capture more important illumination in given time, further improving the robustness of existing light transport simulation methods.}
	\end{center}
\end{figure*}

\section{Results}

We implement our GMIS on a 4-core Intel Core-i7 7700K 4.2GHz processor. All images are rendered progressively with one eye path per pixel at resolution \emph{512*512}. Each iteration starts by tracing the light paths, as the number of light paths is equal to image pixels about 589k. For each light path, we set it could select at most 20 samples in all of its sub-path. At each vertex in a light path, it also have different number of samples, as it may hit on diffuse or specular surface. We compared our GMIS with three well-know light transport simulation algorithms, bidirectional path tracing~\cite{Veach1995}, progressive photon mapping~\cite{Hachisuka:2008:PPM:1409060.1409083}, and vertex connecting and merging~\cite{Georgiev:2012:VCM}. Our GMIS is also east to integrated into current GPU renderer, reaching a apparent speed-up for averages five to ten times faster than CPU implementation.

We have tested four scenes with different characteristics in Figure \textbf{4}. For the \textbf{scene 1}, it contains two metal balls at left and four transparent balls at right part, many ambient light are on the right and back wall. Our GMIS could render more details in the specular ambient shadows. And on the surface of all balls, it contains more details of surrounding objects. BPT fails to handle transparent balls, while PPM also have problems in rendering transparent objects. VCM provides a reasonable results, but in ambient lights and reflectance details, it still has noises. Our GMIS could reduce variance in these areas, like reflectance of surrounding objects on specular surface, the shadows of transparent balls on the wall, it brings more details with less noises.

For the \textbf{scene 2}, it emphasizes the rendering of diffuse objects, as it contains many diffuse surfaces with moderately glossy features, illuminated by one small area light. The pictures show that the GMIS has smooth results in all of diffuse surfaces, while other three methods demonstrate different degrees of white noises in many areas. As the GMIS could have more samples in diffuse surface, the color of these surfaces are smooth since the variance are restrained by increasing number of samples.

The \textbf{scene 3} is for testing the rendering results with mirror. The back wall is a full mirror and the front surface of cube and three balls are all specular as well, which bring in excessive noise. The combination of objects with mirrors may produce caustic paths that BPT could not handle these paths well. PPM and VCM provide similar acceptable results. Our GMIS could handle these paths more robustly, as the reflectance on specular balls are more smooth.

\textbf{Scene 4} contains highly glossy floor with specular chrome features and transparent cube with one area light and one directional light. We can see through the transparent cube and catch sight of two balls behind the cube. This illumination, seen through the
cube, and reflections of surrounding balls are difficult for BPT. Meanwhile, PPM also performs poorly on the glossy floor. Our GMIS could render a more smooth image and more details of balls behind the transparent cube even compared with VCM.

To verify our GMIS has a more speedy convergence rate, we also measure the root mean squared error (RMSE) between the results produced by BPT, PPM and VCM on four test scenes. Figure \textbf{5} shows that a plot of decreasing difference over time. We can see results rendered by GMIS show apparently less RMSE compared with reference images in equal rendering time. This also demonstrate the robustness of GMIS.

\begin{figure}[h]
	\begin{center}
		\includegraphics[height=0.85\textheight]{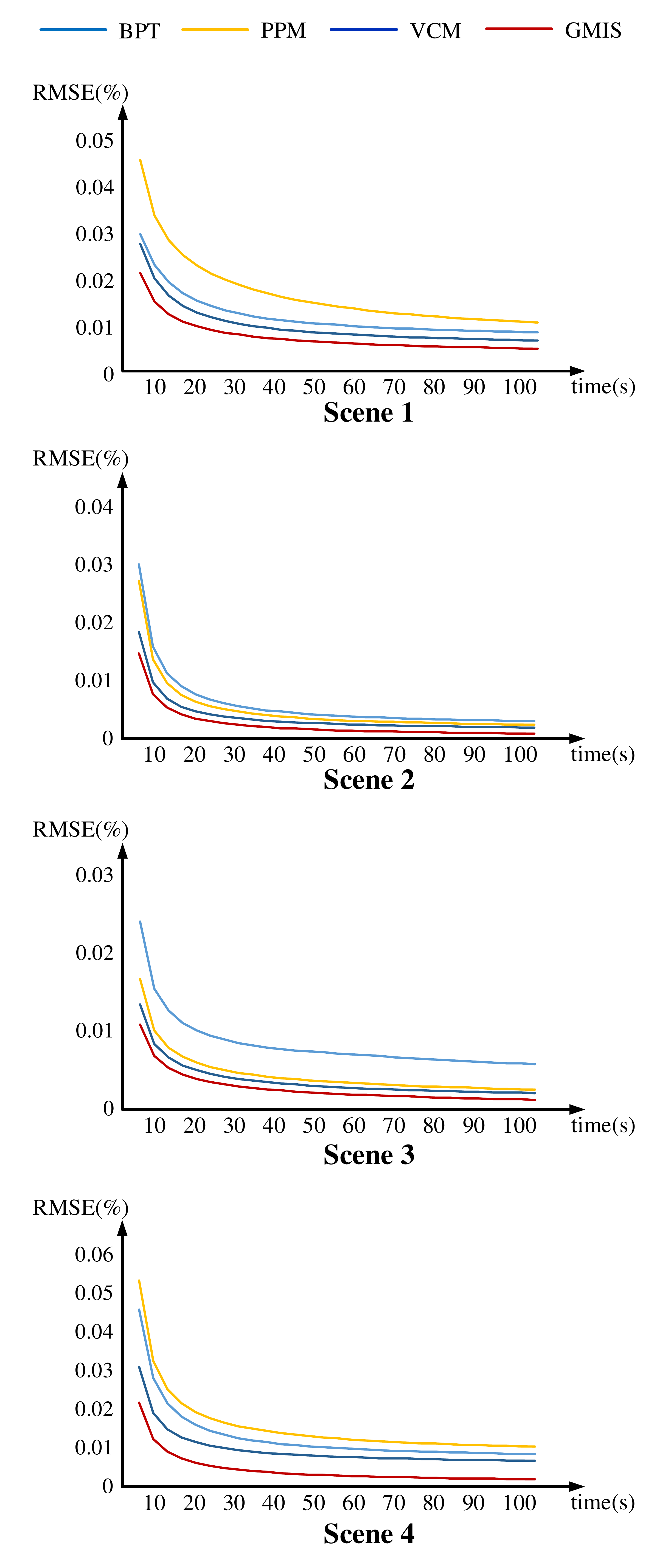}
		\caption{Convergence comparison of GMIS against with BPT, PPM and VCM methods after 100 seconds of rendering. The plots show that our GMIS converges at a higher rate than other three algorithms for different types of scenes.}
	\end{center}
\end{figure}

\section{Conclusions}

Our work presents a more robust numerical integration framework for light transport simulation named generalized multiple importance sampling (GMIS), which combine three sampling strategies and five weighting functions, resulting in fast convergence rate and lower variance. GMIS improve the sampling quality, obtaining more realistic rendering effects with less noises. In the future, we wish to combine this GMIS with learning based light transport algorithms to further improve the performance of photorealistic rendering.


{\small
\bibliographystyle{cvm}
\bibliography{cvmbib}

\begin{thebibliography}{10}\itemsep=-1pt

\bibitem{DBLP:journals/corr/DahmK17}
K.~Dahm and A.~Keller.
\newblock Learning light transport the reinforced way.
\newblock {\em CoRR}, abs/1701.07403, 2017.

\bibitem{Dutré1995}
P.~Dutr{\'e} and Y.~D. Willems.
\newblock {\em Importance-driven Monte Carlo Light Tracing}, pages 188--197.
\newblock Springer Berlin Heidelberg, Berlin, Heidelberg, 1995.

\bibitem{2015arXiv151103095E}
V.~{Elvira}, L.~{Martino}, D.~{Luengo}, and M.~F. {Bugallo}.
\newblock {Generalized Multiple Importance Sampling}.
\newblock {\em ArXiv e-prints}, Nov. 2015.

\bibitem{Georgiev:2012:VCM}
I.~Georgiev, J.~K\v{r}iv\'{a}nek, T.~Davidovi\v{c}, and P.~Slusallek.
\newblock Light transport simulation with vertex connection and merging.
\newblock {\em ACM Trans. Graph.}, 31(6):192:1--192:10, Nov. 2012.

\bibitem{Hachisuka:2009:SPP:1618452.1618487}
T.~Hachisuka and H.~W. Jensen.
\newblock Stochastic progressive photon mapping.
\newblock {\em ACM Trans. Graph.}, 28(5):141:1--141:8, Dec. 2009.

\bibitem{Hachisuka:2014:MML:2601097.2601138}
T.~Hachisuka, A.~S. Kaplanyan, and C.~Dachsbacher.
\newblock Multiplexed metropolis light transport.
\newblock {\em ACM Trans. Graph.}, 33(4):100:1--100:10, July 2014.

\bibitem{Hachisuka:2008:PPM:1409060.1409083}
T.~Hachisuka, S.~Ogaki, and H.~W. Jensen.
\newblock Progressive photon mapping.
\newblock {\em ACM Trans. Graph.}, 27(5):130:1--130:8, Dec. 2008.

\bibitem{Hachisuka:2012:PSE:2366145.2366210}
T.~Hachisuka, J.~Pantaleoni, and H.~W. Jensen.
\newblock A path space extension for robust light transport simulation.
\newblock {\em ACM Trans. Graph.}, 31(6):191:1--191:10, Nov. 2012.

\bibitem{CGF:CGF12950}
S.~Herholz, O.~Elek, J.~Vorba, H.~Lensch, and J.~Křivánek.
\newblock Product importance sampling for light transport path guiding.
\newblock {\em Computer Graphics Forum}, 35(4):67--77, 2016.

\bibitem{Jensen:2001:RIS:500844}
H.~W. Jensen.
\newblock {\em Realistic Image Synthesis Using Photon Mapping}.
\newblock A. K. Peters, Ltd., Natick, MA, USA, 2001.

\bibitem{Kajiya:1986:RE:15886.15902}
J.~T. Kajiya.
\newblock The rendering equation.
\newblock {\em SIGGRAPH Comput. Graph.}, 20(4):143--150, Aug. 1986.

\bibitem{Kalantari:2015:MLA:2809654.2766977}
N.~K. Kalantari, S.~Bako, and P.~Sen.
\newblock A machine learning approach for filtering monte carlo noise.
\newblock {\em ACM Trans. Graph.}, 34(4):122:1--122:12, July 2015.

\bibitem{CGF:CGF703}
C.~Kelemen, L.~Szirmay-Kalos, G.~Antal, and F.~Csonka.
\newblock A simple and robust mutation strategy for the metropolis light
  transport algorithm.
\newblock {\em Computer Graphics Forum}, 21(3):531--540, 2002.

\bibitem{Lafortune93bi-directionalpath}
E.~P. Lafortune and Y.~D. Willems.
\newblock Bi-directional path tracing.
\newblock In {\em PROCEEDINGS OF THIRD INTERNATIONAL CONFERENCE ON
  COMPUTATIONAL GRAPHICS AND VISUALIZATION TECHNIQUES (COMPUGRAPHICS ’93},
  pages 145--153, 1993.

\bibitem{10.1111:cgf.13227}
T.~Müller, M.~Gross, and J.~Novák.
\newblock {Practical Path Guiding for Efficient Light-transport Simulation}.
\newblock {\em Computer Graphics Forum}, 2017.

\bibitem{Nalbach:2017:DSC:3128450.3128458}
O.~Nalbach, E.~Arabadzhiyska, D.~Mehta, H.-P. Seidel, and T.~Ritschel.
\newblock Deep shading: Convolutional neural networks for screen space shading.
\newblock {\em Comput. Graph. Forum}, 36(4):65--78, July 2017.

\bibitem{article}
F.~O.~Bartell, E.~Dereniak, and W.~L.~Wolfe.
\newblock The theory and measurement of bidirectional reflectance distribution
  function (brdf) and bidirectional transmittance distribution function (btdf).
\newblock 257:154--160, 01 1980.

\bibitem{Pajot:2011:RRA:1999161.1999196}
A.~Pajot, L.~Barthe, M.~Paulin, and P.~Poulin.
\newblock Representativity for robust and adaptive multiple importance
  sampling.
\newblock {\em IEEE Transactions on Visualization and Computer Graphics},
  17(8):1108--1121, Aug. 2011.

\bibitem{Veach:1998:RMC:927297}
E.~Veach.
\newblock {\em Robust Monte Carlo Methods for Light Transport Simulation}.
\newblock PhD thesis, Stanford, CA, USA, 1998.
\newblock AAI9837162.

\bibitem{Veach1995}
E.~Veach and L.~Guibas.
\newblock {\em Bidirectional Estimators for Light Transport}, pages 145--167.
\newblock Springer Berlin Heidelberg, Berlin, Heidelberg, 1995.

\bibitem{Veach:1995:OCS:218380.218498}
E.~Veach and L.~J. Guibas.
\newblock Optimally combining sampling techniques for monte carlo rendering.
\newblock In {\em Proceedings of the 22Nd Annual Conference on Computer
  Graphics and Interactive Techniques}, SIGGRAPH '95, pages 419--428, New York,
  NY, USA, 1995. ACM.

\bibitem{Veach:1997:MLT:258734.258775}
E.~Veach and L.~J. Guibas.
\newblock Metropolis light transport.
\newblock In {\em Proceedings of the 24th Annual Conference on Computer
  Graphics and Interactive Techniques}, SIGGRAPH '97, pages 65--76, New York,
  NY, USA, 1997. ACM Press/Addison-Wesley Publishing Co.

\bibitem{Vorba:2014:OLP:2601097.2601203}
J.~Vorba, O.~Karl\'{\i}k, M.~\v{S}ik, T.~Ritschel, and J.~K\v{r}iv\'{a}nek.
\newblock On-line learning of parametric mixture models for light transport
  simulation.
\newblock {\em ACM Trans. Graph.}, 33(4):101:1--101:11, July 2014.

\end{thebibliography}
}

\end{document}